\documentclass[letterpaper, 10 pt, conference]{ieeeconf} 
\IEEEoverridecommandlockouts 
\usepackage{amsmath,amsfonts}
\usepackage{amssymb}
\usepackage{amsthm}
\usepackage{array}
\usepackage{graphicx}
\usepackage{textcomp}
\usepackage[justification=justified, font=small]{caption}
\usepackage{color}
\usepackage{url}
\usepackage{verbatim}
\usepackage{algorithm}
\usepackage{subfigure}
\usepackage{algpseudocode}
\usepackage{tikz}
\usepackage{xcolor}
\usepackage{mdframed}
\usepackage{hyperref}
\usepackage{balance}
\usepackage{cite}
\usepackage{flushend}
\usepackage{soul}
\usepackage{courier}
\usepackage{epstopdf}
\newtheorem{definition}{Definition}

\newtheorem{remark}{Remark}

\newcommand{\AP}{\mathsf{AP}}

\title{Risk-Aware Autonomous Driving with Linear Temporal Logic Specifications}
\author{Shuhao Qi$^{1}$ , Zengjie Zhang$^{1}$ , Zhiyong Sun$^{2}$ and Sofie Haesaert$^{1}$ 
\thanks{This work is supported by the European project SymAware under grant No. 101070802, the European project COVER under grant No. 101086228, and the Dutch NWO Veni project CODEC under grant No. 18244. Corresponding author: \textit{Zhiyong Sun}. }
\thanks{$^{1}$ S. Qi, Z. Zhang, and S. Haesaert are with the Department of Electrical Engineering, Eindhoven University of Technology, Eindhoven, The Netherlands.
        {\tt\small \{s.qi, z.zhang3, s.haesaert\}@tue.nl}}
\thanks{$^{2}$Z. Sun is with the College of Engineering, Peking University, Beijing, China. {\tt\small \{zhiyong.sun@pku.edu.cn\}}}
}

\begin{document}
\maketitle

\begin{abstract} 
Human drivers naturally balance the risks of different concerns while driving, including traffic rule violations, minor accidents, and fatalities. However, achieving the same behavior in autonomous driving systems remains an open problem. This paper extends a risk metric that has been verified in human-like driving studies to encompass more complex driving scenarios specified by linear temporal logic (LTL) that go beyond just collision risks. This extension incorporates the timing and severity of events into LTL specifications, thereby reflecting a human-like risk awareness. Without sacrificing expressivity for traffic rules, we adopt LTL specifications composed of safety and co-safety formulas, allowing the control synthesis problem to be reformulated as a reachability problem. By leveraging occupation measures, we further formulate a linear programming (LP) problem for this LTL-based risk metric. Consequently, the synthesized policy balances different types of driving risks, including both collision risks and traffic rule violations. The effectiveness of the proposed approach is validated by three typical traffic scenarios in \textit{Carla} simulator.
\end{abstract}
\section{Introduction}

\begin{figure}[htbp]
    \centering
    \includegraphics[width=0.45
    \textwidth]{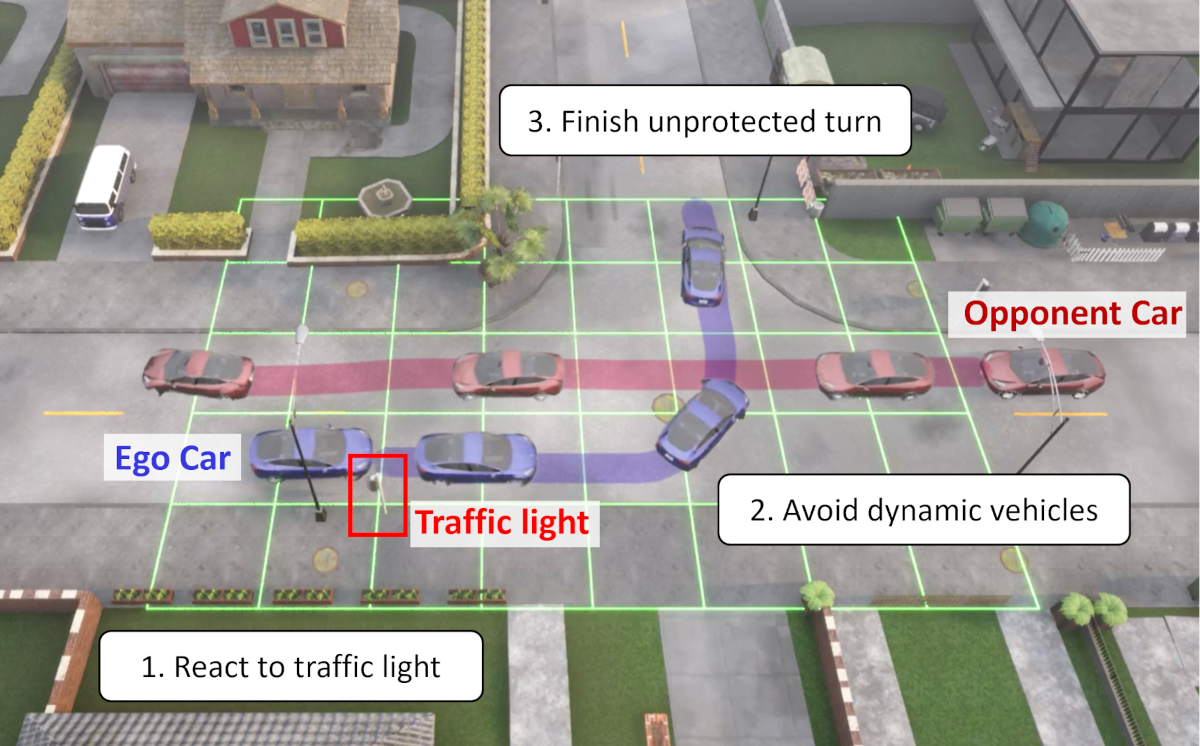}
    \caption{Unprotected turn in an intersection}
    \label{fig:intersection}
\end{figure}

Developing motion planning frameworks for autonomous vehicles remains a significant challenge due to the inherent complexity and uncertainty of real-world traffic~\cite{mehdipour2023formal}. The complexity arises primarily from the diverse concerns in practical driving tasks, including safety, traffic rules, and social norms~\cite{wang2022social}. The uncertainty mainly originates from the dynamic environment with unpredictable traffic participants and unexpected events~\cite{wongpiromsarn2021minimum}. Taking an unprotected turn scenario in Fig.~\ref{fig:intersection} as an example, the ego vehicle must make a left turn while avoiding collisions with the opponent car, obey traffic rules such as reacting to traffic lights, and follow driving norms like yielding to oncoming vehicles. Environmental uncertainty places the ego vehicle at the potential risk of violating the above specifications. Therefore, a risk-aware autonomous driving scheme is essential to make decisions that balance diverse risks associated with safety, traffic rules, and social norms. 

Temporal logic is a formal and expressive language used to describe both spatial and temporal properties of systems through logical and temporal operators~\cite{belta2017formal}. It has been widely applied in verifying software and digital circuits, and more recently explored in motion planning of autonomous vehicles, as it is well-suited for encoding high-level traffic rules~\cite{althoff2025no} and synthesizing provably correct controllers~\cite{wongpiromsarn2023formal, qi2023automated}. However, current temporal logic-based planning frameworks remain limited and impractical for traffic applications due to inherent uncertainty in the real world. Conventional temporal logic planning aims to achieve complete safety and full satisfaction of temporal logic specifications~\cite{qi2023automated}, which is unrealistic in practice. To handle uncertainty, existing methods quantify risk using the probability that a temporal logic specification fails and focus on maximizing the satisfaction probability of these specifications~\cite{guo2018probabilistic, lindemann2021stl, van2022temporal}. However, such risk metrics are not suitable for realistic traffic where different types of risky events are involved~\cite{lindemann2021reactive}. In real-world traffic with diverse risks, human drivers intuitively prioritize near-future and more severe events over those in the distant future or with less severity~\cite{geisslinger2021autonomous}. Unfortunately, the satisfaction probability of temporal logic fails to differentiate the severity and timing of multiple events, which may result in illogical and even unsafe behavior.

To achieve human-like risk assessment, the driver's risk field model is proposed in the transportation domain, leveraging an expert-designed distribution to emulate human drivers’ awareness of predictive events~\cite{kolekar2020human, joo2023generalized}. This heuristic field model enables vehicles to differentiate the timing and severity of potential collisions and track reference paths safely by balancing risks among multiple obstacles. However, this model is limited to collision avoidance in tracking tasks, making it less effective for handling realistic traffic scenarios with complex rules. To address this, we extend the risk field model to temporal logic specifications. This paper uses linear temporal logic (LTL)~\cite{belta2017formal}, a classic type of temporal logic widely used in motion planning. By incorporating the inherent ability to balance timing and severity from the risk field model, this extension overcomes the limitation of temporal logic mentioned above. Specifically, we extend upon the tracking task with more complex driving goals expressed by co-safe specifications and quantify risk specified as violating safety specifications. Safety or co-safety specifications~\cite{artale2023complexity} are special fragments of LTL that can transform the control synthesis problem into a classical reachability problem. Therefore, this approach reduces the LTL planning into a reachability problem while preserving the necessary expressivity for traffic tasks. To formulate a tractable control synthesis problem, we rewrite the risk metric using the discounted version of occupation measures, a key concept used in the stochastic reachability problem for constrained Markov Decision Process (MDP) models~\cite{altman2021constrained}. Referring to the multi-objective control synthesis formulation for LTL in~\cite{haesaert2021formal}, we formulate a linear programming (LP) problem over a product automaton to synthesize risk-aware policies using occupation measures as decision variables.

The main contributions of this paper are the introduction of a human-like LTL-based risk metric and the development of a corresponding risk-aware controller to balance different types of risks. The effectiveness and scalability of the proposed method are validated in three distinct driving scenarios on Carla simulator~\cite{dosovitskiy2017carla}, demonstrating its capability to handle complexities from traffic rules and unexpected events due to uncertainty, and its scalability across versatile scenarios.

\section{Preliminary and Problem Statement}\label{sec:form}


\subsection{Finite-State Model for Vehicles and Environments}

MDP is a typical mathematical framework for decision-making in stochastic environments, and its variants, such as the partially observable MDP, can characterize more complex traffic scenarios~\cite{kochenderfer2015decision}. As the first step before considering more complex models, this paper begins by using a finite-state MDP to capture the uncertain behavior of vehicles.
\begin{definition}
A Markov decision process (MDP) is a tuple $\mathcal{M}=(S, s_0, A, \mathbb{T})$ with $S$ is a finite set of states and $s_0 \in S$ is an initial state; $A$ is a finite set of actions; and $\mathbb{T}\!: S \times\!A\!\times\!S\!\rightarrow [0, 1]$ is a probabilistic transition kernel.\qed
\label{def:mdp}
\end{definition}
We denote a finite path of $\mathcal{M}$ as $\pmb{s}_n\!:=\!s_0 a_0 \ldots a_{n-1} s_n$, $n\!\in\!\mathbb{N}$, where $s_i\!\in\!S$ and $a_i\!\in\!A$ for all $i\!\leq\!n$. 
Similarly, we denote an infinite paths as
$\mathbf{s}\!=\!s_0 a_0 s_1 a_1 \ldots$. The inputs of $\mathcal{M}$ are selected according to a policy defined as follows.


\begin{definition}
A policy is a sequence $\pi\!:=\!\pi_0 \pi_1 \pi_2 \ldots$ of   
stochastic kernels $\pi_n\!:\!H \!\times\!A\!\rightarrow\![0, 1]$, where $H$ denotes the collection of all the finite paths $\pmb{s}_n$. The set of all policies is denoted by $\Pi$.
\label{def:policy}
\end{definition}

\noindent Given a policy $\pi\!\in\!\Pi$ and a finite path $\pmb{s}_n$, the distribution of the next control input $a_n$ is given by $\pi_n\left(\cdot\!\mid\! \pmb{s}_n\right)$ with $\sum_{a_n}\!\pi_n\left(a_n \mid \pmb{s}_n\right)\!=\!1$. 
Given the initial state $s_0$, the stochastic kernel $s_{n+1}\!\sim\!\mathbb{T}\left(\cdot \mid s_n, a_n\right)$ together with the policy $a_n\!\sim\!\pi_n\left(\cdot \mid \pmb{s}_n\right)$ induce a unique probability distribution over the paths of $\mathcal{M}$, which we denote by $\mathbb{P}_\pi$. Policies that depend on $\pmb{s}_n$ only through the current state $s_n$ are called \textit{Markov policies}, which means $\pi_n\!:\!S\!\times\!A\!\rightarrow\![0, 1]$. A Markov policy is called stationary if the kernels $\pi_n$ do not depend on the time index $n$.  When it is applied to the MDP, the resulting model is called a Markov Chain (MC), defined as follows,


\begin{definition}\label{def:mc}
A Markov Chain (MC) is defined as a tuple $\mathcal{C}=(S, s_0, \mathbb{T})$
where $S$ and $s_0$ are defined as in the MDP, and $\mathbb{T}\!:\!S\!\times\!S\!\rightarrow\![0,1]$ is a probabilistic transition kernel. \qed
\end{definition}
Since environmental states are uncontrollable, an MC is used to represent the evolution of environments~\cite{ulusoy2014incremental}. In summary, an MDP, denoted by $\mathcal{M}_v\!=\!(S_v, s_{v}^0, A, \mathbb{T}_v)$, is used to represent the behavior of the ego vehicle, while an MC model, denoted by $\mathcal{C}_e \!=\!(S_e, s_{e}^0, \mathbb{T}_e)$, captures the behavior of the environment. Furthermore, the composition of $\mathcal{M}_v$ and $\mathcal{C}_e$ forms a new MDP model, defined as $\overline{\mathcal{M}}=(\bar{S}, \bar{s}_{0}, A, \bar{\mathbb{T}})$, where $\bar{S}\!:=\!S_v\!\times\!S_e$, $\bar{s}_0\!=\!(s_{v}^0, s_{e}^0)$, and $\bar{\mathbb{T}}\!:\!\bar{S}\!\times\!A\!\times\!\bar{S}\!\rightarrow\![0,1]$ based on $\bar{\mathbb{T}}(\bar{s}, a, \bar{s}^\prime) = \mathbb{T}_v(s_v, a, s_v^\prime) \times \mathbb{T}_e(s_e, s_e^\prime)$.

\subsection{Specifications: Linear Temporal Logic (LTL)}\label{sec:ltl}
The set of atomic propositions $\AP\!=\!\left\{ p_1, \dots, p_N \right\}$ defines an alphabet $\Sigma\!:=\!2^{\AP}$, where each letter $\omega\!\in\!\Sigma$ contains the set of atomic propositions that are true. A labeling function $\mathcal{L}$ maps states to letters in the alphabet $\Sigma$, such that an infinite path $\pmb{s}\!=\!s_0 s_1 s_2 \dots$ generates a word as
$\pmb{\omega}\!:=\!\mathcal{L}(s_0) \mathcal{L}(s_1) \mathcal{L}(s_2)\dots$. A suffix of a word
$\pmb{\omega}$ is $\pmb{\omega}_k\!=\!\omega_k \omega_{k+1} \omega_{k+2} \dots$, where $\omega_i \in \Sigma$ and $k\!\in\!\mathbb{N}_{\geq 0}$. 

\noindent \textbf{Syntax:} An LTL~\cite{belta2017formal} specification is recursively defined as,
\begin{equation}
    \psi ::= \top \mid p \mid \neg \psi \mid \psi_1 \wedge \psi_2 \mid \bigcirc \psi \mid \psi_1 \mathsf{U} \psi_2,
\end{equation}
where  $\psi_1$, $\psi_2$ and $\psi$ are LTL formulas, $p \in \AP$ is an atomic proposition, $\neg$ is the \textit{negation} operator, $\wedge$ is the \textit{conjunction} operator that connects two LTL formulas, and $\bigcirc$ and $\mathsf{U}$ represent the \textit{next} and \textit{until} temporal operators, respectively. Other logical and temporal operators, namely \textit{disjunction} $\vee$, \textit{implication} $\rightarrow$, \textit{eventually} $\lozenge$, and \textit{always} $\square$ can be defined as, $\psi_1 \vee \psi_2\!:=\!\lnot \!\left(\lnot\psi_1 \wedge \lnot\psi_2 \right)$, $\psi_1\!\rightarrow\!\psi_2\!:=\!\lnot \psi_1 \vee \psi_2$, $\lozenge \psi\!:=\!\top \mathsf{U} \psi$, and $\square \psi \!:=\!\neg \lozenge \neg \psi$. 


\noindent{\textbf{Semantics:}} For a given word $\pmb{\omega}$, basic semantics of LTL are given as
$\pmb\omega_k \!\models\! p$, if $p \in \omega_k$; $\pmb\omega_k \!\models\! \lnot p$, if $p \notin \omega_k$;
$\pmb\omega_k \!\models\! \psi_1 \wedge \psi_2$, if $\pmb \omega_k \!\models\! \psi_1$ and $\pmb \omega_k \!\models\! \psi_2$;
$\pmb\omega_k \!\models\! \bigcirc \psi$, if $\pmb\omega_{k+1} \!\models\! \psi$;
$\pmb\omega_k \!\models\! \psi_1 \mathsf{U} \psi_2$, if $\exists$ $i \in \mathbb{N}$ such that $\pmb\omega_{k+i} \!\models\! \psi_2$, and $\pmb\omega_{k+j}\!\models\!\psi_1$ holds $\forall \, 0\!\leq j \!< \!i$.

\noindent \textbf{Fragments:} Two notable fragments of LTL are the safety and co-safety specifications~\cite{artale2023complexity}, which allows reasoning about the satisfaction of an infinite word using only a finite prefix. Specifically, a co-safety formula is satisfied if there exists a finite ``good" prefix regardless of any infinite suffix. Conversely, the violation of a safety specification depends on the existence of a finite bad prefix. In other words, safety fragments represent ``bad thing never happens", while co-safety fragments depict ``good thing eventually happens". A syntactically co-safety LTL (scLTL) formula $\psi$ over a set of atomic propositions $\AP$ is recursively defined as,
$$
\psi::= \text{true} \mid p \mid \neg p \mid \psi_1 \wedge \psi_2\left|\psi_1 \vee \psi_2\right| \bigcirc \psi \mid \psi_1 \mathsf{U} \psi_2
$$
where $\psi_1$, $\psi_2$, and $\psi$ are scLTL formulas, and $p \in \mathrm{AP}$. An LTL formula, $\psi$, is a safety specification if and only if its negation $\neg \psi$ is a co-safety specification.



\noindent \textbf{Automaton:} A general LTL specification can be translated into a B\"{u}chi automaton~\cite{belta2017formal} whose semantics are defined over infinite words. In contrast, scLTL and safety specification can be also translated into deterministic finite-state automatons (DFA), defined as follows:
\begin{definition}\label{def:dfa} 
An DFA is a tuple $\mathcal{A}=\left(Q, q_0, \Sigma, \delta, F\right)$, where $Q$ is a finite set of states and $q_0 \in Q$ is an initial state; $\Sigma$ is a finite input alphabet; $\delta: Q \times \Sigma \rightarrow Q$ is a deterministic transition function; $F \subset Q$ is a set of final states. \qed
\end{definition}
\noindent Final states are sink states, i.e., a path entering it stays there forever. The final states in the DFA from a co-safety specification are accepting states, while the final states in the DFA of a safety specification are non-accepting. 


\noindent \textbf{Study Case:} In an intersection scenario depicted in Fig.~\ref{fig:intersection}, the ego vehicle is expected to eventually reach a target area labeled ``$t$", avoid collisions with the other vehicle labeled ``$v$", refrain from entering non-drivable areas labeled ``$n$", and also wait for a green light event labeled ``$g$" before entering the intersection area labeled ``$i$". In this scenario, the LTL specification is $\psi =  \psi_{cs} \wedge \psi_{s}$ with a co-safety formula $\psi_{cs} = \lozenge(t)$ and a safety formula $\psi_{s} = \square(\neg g\!\rightarrow\!\neg i) \wedge \square(\neg n \wedge \neg v)$, defined over $\AP\!=\!\{t, v, g, i\}$. DFAs can be autonomously translated from $\psi_s$ and $\psi_{cs}$ using off-the-shelf toolboxes~\cite{henriksen1995mona}.

\subsection{Risk-Aware Control Synthesis Problem}
\label{sec:problem}

Since traffic rules are primarily concerned with the occurrence of good or bad events rather than repetitive behaviors,  we use the combination of safety and co-safety fragments to effectively represent nearly all traffic tasks and rules while mitigating the computational demand of control synthesis. Thus, we consider an LTL specification in the form of $\psi = \psi_{cs} \wedge \psi_s$ defined over an atomic proposition set $\AP$, composed with a safety formula $\psi_s$ and a co-safety formula $\psi_{cs}$. The corresponding DFAs of $\psi_{cs}$ and $\psi_{s}$ are denoted by $\mathcal{A}_{cs}$
and $\mathcal{A}_{s}$, respectively. To represent the uncertain behavior of both the environment and vehicle, we use a composed MDP $\overline{\mathcal{M}}$, as defined in Def.~\ref{def:mdp}. This composed MDP combines an MDP model for an ego vehicle, $\mathcal{M}_v$, and an MC model for an uncontrollable environment, $\mathcal{C}_e$, defined in Def.~\ref{def:mc}. 
The states of $\overline{\mathcal{M}}$ are labeled by $\mathcal{L}$ to an alphabet set $\Sigma$, which can be used to verify the satisfaction of $\psi$. To handle the uncertainty in realistic traffic like a human driver, we aim to develop a human-like risk metric for $\psi$ and then synthesize a policy $\pi\!\in\!\Pi$ that balances different types of risk and drives the ego vehicle to satisfy $\psi$.

\section{human-like risk metric}\label{sec:risk}
Risk assessments are often discussed on simple reach-avoid problems~\cite{majumdar2020should}. Therefore, we start by rewriting the satisfaction problem for the given specifications $\psi = \psi_{cs} \wedge \psi_s$ as a reach-avoid problem over product MDP.

\subsection{Product MDP and Reachability Problem}

\begin{definition}\label{def:product}
    The product MDP of $\overline{\mathcal{M}}$, $\mathcal{A}_{cs}$ and $\mathcal{A}_s$ is defined as a tuple $\mathcal{P}\!=\!(Z, z_0, A, \hat{\mathbb{T}}, G, D)$, with the state set $Z\!:=\!S\!\times\!Q_{cs}\!\times\!Q_{s}$, the action set $A$, and the initial state $z_0\!=\!(\bar{s}_0, \delta_{cs}(q_{cs}^0, \mathcal{L}(\bar{s}_0)), \delta_{s}(q_s^0, \mathcal{L}\left(\bar{s}_0))\right.$. $\hat{\mathbb{T}}\!:\!Z\!\times\!A\!\times\!Z\!\rightarrow \![0,1]$ is the probabilistic transition kernel, which can specify a transition from state $z_n\!=\!(\bar{s}_n, q_s^n, q_{cs}^n)$ to $z_{n+1}\!=\!(\bar{s}_{n+1}, q_s^{n+1}, q_{cs}^{n+1})$ under action $a_n\!\in\!A$ based on $\bar{s}_{n+1}\!\sim\!\bar{\mathbb{T}}(\cdot\!\mid\!\bar{s}_n, a_n)$, $q_{cs}^{n+1}\!=\!\delta_{cs}(q_{cs}^n, \mathcal{L}(\bar{s}_n))$ and $q_s^{n+1}\!=\!\delta_{s}(q_s^n, \mathcal{L}(\bar{s}_n))$ for each $n \in \mathbb{N}$. Denote the accepting set for $\psi_{cs}$ as $G\!:=\!\bar{S}\!\times\!F_{cs}$, and the non-accepting set for $\psi_{s}$ as $D\!:=\!\bar{S}\!\times\!F_{s}$.
\end{definition}
It is well known from~\cite{belta2017formal} that the satisfaction probability of a co-safety specification equates to a reachability probability towards the accepting state set of the product automaton defined in Def.~\ref{def:product}, 
\begin{equation}\label{eq:cosafe_prob}
\textstyle    \mathbb{P}_\pi(\pmb{\omega} \vDash \psi_{cs})=\overline{\mathbb{P}}_{\bar{\pi}}\left(\lozenge G \right) =  \mathbb{E}_{\bar{\pi}} \left[\sum_t^\infty \mathbf{1}(z_t\!\in\!G) \right],
\end{equation}
\noindent where $\mathbf{1}(\cdot)$ is an indicator function, $\bar{\pi}$ is policy for $\mathcal{P}$, and $\overline{\mathbb{P}}_\pi$ denotes the probability distribution over the paths of $\mathcal{P}$ under $\bar{\pi}$. Note that $\bar{\pi}$ for $\mathcal{P}$ is a Markovian and stationary policy, which is converted from $\pi$ for $\mathcal{M}_v$ defined in Def.~\ref{def:policy}. Similarly, for a safety specification $\psi_s$, the violation probability equates to a reachability probability towards the non-accepting state set of $\mathcal{P}$,
\begin{equation}\label{eq:unsafe_prob}
\textstyle    \mathbb{P}_\pi(\pmb{\omega} \nvDash \psi_s)=\overline{\mathbb{P}}_{\bar{\pi}}\left(\lozenge D \right) = \mathbb{E}_{\bar{\pi}} \left[\sum_t^\infty \mathbf{1}(z_t\!\in\!D) \right].
\end{equation}

\subsection{Limitations of Event Probability}
\begin{figure}[htb]
    \centering
    \vspace{-2mm}
    \includegraphics[width=0.25
    \textwidth]{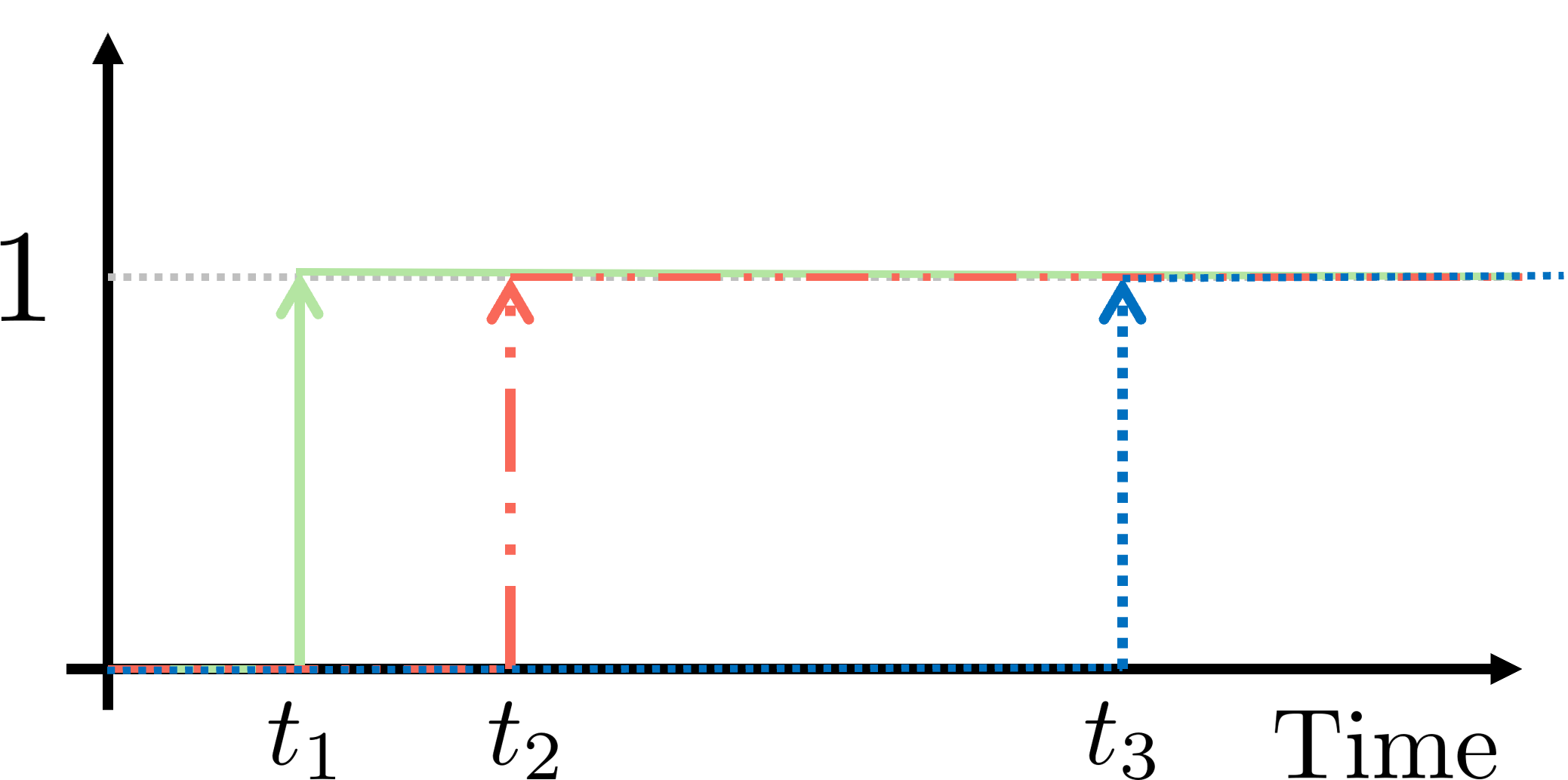}
    \caption{Sketch for $\mathbb{P}_\pi$ with three traces in different colors. }
    \label{fig:probability} \vspace{-3mm}
\end{figure}

The event probability in~\eqref{eq:cosafe_prob} and~\eqref{eq:unsafe_prob} means that each possible trace generated by $\pi$ gets weight 1 if it enters into final states, and otherwise gets weighted 0. As displayed in Fig.~\ref{fig:probability}, three different traces are considered identical, even when they enter the final set at different times. Such an approach embraces two points against the awareness way of human drivers: (1) It does not account for the timing of violations, which may cause the vehicle to react too early to dangers that happen in the distant future; (2) It fails to distinguish different types of events, preventing the vehicle from appropriately account for the severity of multiple risks. Thus, a new risk metric for LTL specifications reflecting human-like awareness is necessary.

\subsection{Human-like Metric Design}
Discounted rewards are commonly used in reinforcement learning and economics to account for the timing of events when decisions lead to rewards at different future moments. The satisfaction probability, $\mathbb{P}_\pi (\pmb{\omega} \models \psi_{cs})$, can be modified by introducing discounting as follows,
\begin{equation}
\textstyle \mathcal{V}_{\bar{\pi}} = \mathbb{E}_{\bar{\pi}} \left[\sum_t^\infty \gamma^t \mathbf{1}(z_t \in G)\right],
 \label{eq:reward_metric}
\end{equation}
where $\gamma \!<\! 1$ is a discount factor. This discounting represents the preference of people for obtaining rewards sooner rather than later. The rationality of the discounting operation has also been supported by psychology studies~\cite{ahmed2020rationality}. When it comes to violation risks of safety specifications, we propose a similar metric. In addition, varying severity of events is considered by a cost mapping function, denoted as $c(\cdot)\!:\!Z\!\rightarrow\!\mathbb{R}$, that assigns cost values to different product states. Specifically, $c(\hat{z})= 0$ if $\hat{z}\!\notin\!D$, while $c(\hat{z})> 0$ if $\hat{z}\!\in\!D$. Furthermore, we propose the following metric to evaluate the risk of a policy $\pi$,
\begin{equation}
\textstyle \mathcal{R}_{\bar{\pi}} = \mathbb{E}_{\bar{\pi}} \left[\sum_t^\infty \gamma^t c(z_t)\right].
 \label{eq:risk_metric}
\end{equation}

\begin{figure}[tb]
\vspace{1mm}
    \centering
    \includegraphics[width=0.22
    \textwidth]{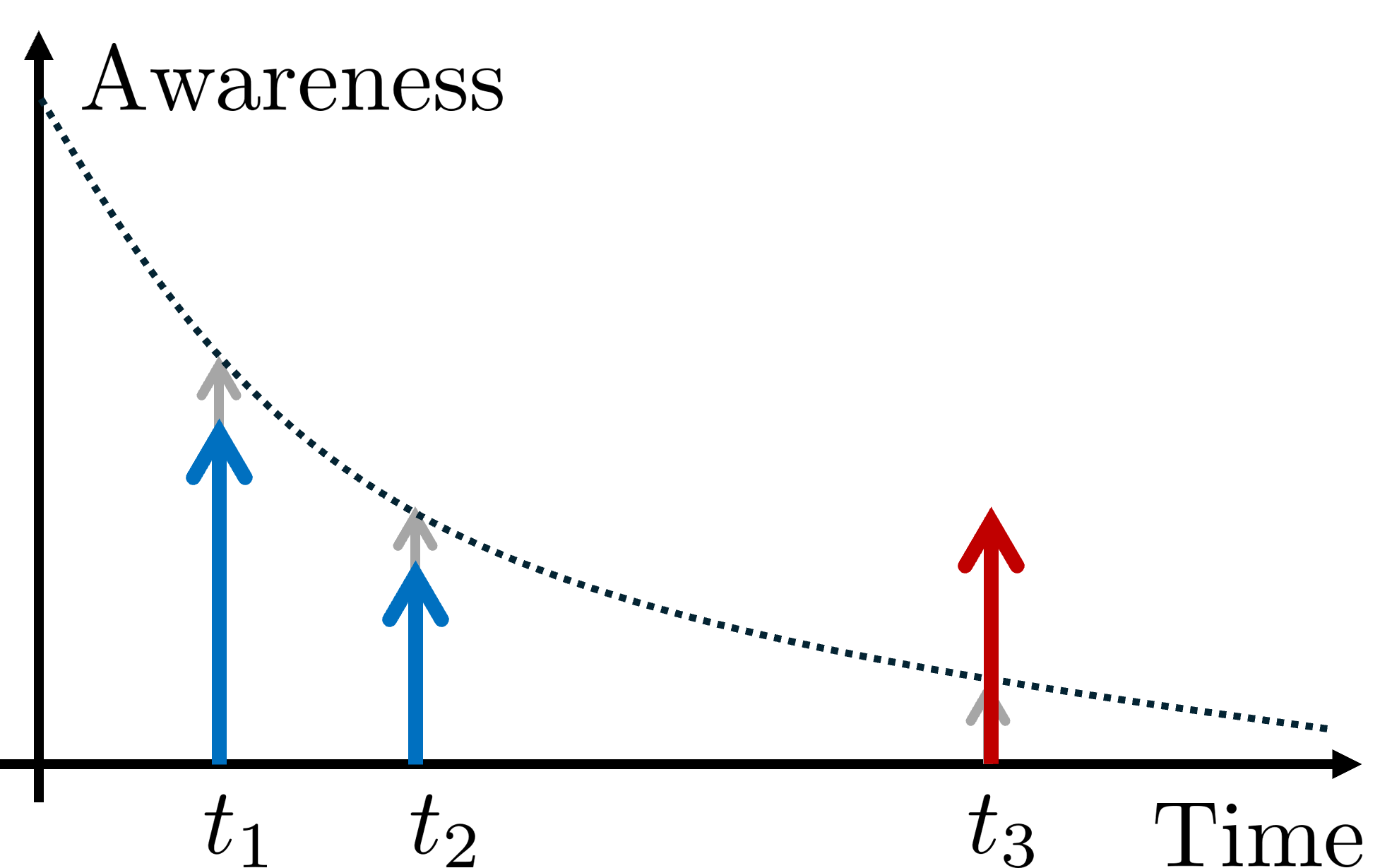}
    \caption{Sketch for $\mathcal{R}_{\bar{\pi}}$. Red and blue indicate two types of events with different severity levels. The grey arrows illustrate decreasing awareness levels over the horizon due to discounting, while the colored arrows represent the corresponding risk values.}
    \label{fig:risk}
\end{figure}

\noindent The proposed metric addresses the limitations of event probabilities. As depicted in Fig.~\ref{fig:risk}, discounting reflects how a driver's awareness gradually decreases over time, causing the perceived risk for the same event to decrease accordingly. When an event is significantly severe, like the one indicated by the red arrow, the risk metric in~\eqref{eq:risk_metric} places greater importance on this event than on less severe ones. Therefore, this metric can account for both the timing and severity of risky events. Defined over product state space, this metric can be seen as an extension of the driver’s risk field (DRF) model~\cite{kolekar2020human, joo2023generalized} to LTL specifications.

\begin{mdframed} 
\itshape
\textbf{Driver’s Risk Field (DRF)} is a heuristic model to evaluate the risks from a driver's view. The risk field $\alpha(\cdot)\!:\!\mathbb{R}^2\!\rightarrow\!\mathbb R$ maps planar coordinates to scalar values representing the driver's attention. To model human-like awareness, DRF is designed based on the intuition that the importance of future events decreases while the corresponding uncertainty increases along the predictive horizon. A detailed formulation of DRF can be found in~\cite{kolekar2020human}. Perceived risk is evaluated by calculating the expected event cost, which is the product of the risk field and the cost map. With a grid discretization, the DRF risk metric is defined as:
 \begin{equation}
\textstyle \mathcal{R} = \sum_{x \in X} \sum_{y \in Y} c(x, y) \alpha(x, y),
\label{eq:drf_risk}
\end{equation}
\noindent where $c(\cdot)\!:\!\mathbb{R}^2\!\rightarrow \!\mathbb R$ represents the cost distribution over planar coordinates, and $X$ and $Y$ are finite sets of discretized coordinates along with $x$ and $y$-axis.
\end{mdframed}

Inspired by DRF risk metric in Eq.~\eqref{eq:drf_risk} , the risk metric in Eq.~\eqref{eq:risk_metric} is rewritten as follows: 
\begin{equation}
    \begin{aligned}
        \mathcal{R}_{\bar{\pi}} & \textstyle = \mathbb{E}_{\bar{\pi}} \left[\sum_t^\infty \sum_{z\in Z}  \gamma^t c(z) \mathbf{1} (z_t = z)\right]  \\
        & \textstyle = \sum_{z\in Z} c(z)  \underbrace {\mathbb{E}_{\bar{\pi}} \textstyle \left[ \sum_t^\infty \gamma^t  \mathbf{1} (z_t = z) \right]}_{\beta_{\bar{\pi}}(z)}.
    \end{aligned}
    \label{eq:oc_risk_metric}
\end{equation}

\noindent Referring to the standard occupation measure defined as $\mathbb{E}_{\bar{\pi}} \textstyle \left[ \sum_t^\infty \mathbf{1} (z_t = z) \right]$, $\beta_{\bar{\pi}}(z)$ can be regarded as a discounted version of the occupation measure. An occupation measure describes the expected number of times that a state has been visited before reaching the final set, given a Markov and stationary policy $\bar{\pi}(z, a)$. Occupation measures can be used to formulate stochastic shortest path problems~\cite{trevizan2017occupation}. By comparing risk metrics in~\eqref{eq:oc_risk_metric} and~\eqref{eq:drf_risk}, it is evident that the discounted occupation measure plays a similar role as the DRF model. In addition, the discounted occupation measure embraces similar awareness for events as the DRF model. Thus, we conjecture that the proposed risk metric captures a similar human-like risk awareness as DRF, which will be validated by subsequent simulation results. 

\begin{remark} The cost function is designed by experts in this work but can potentially be learned from real-world driving data to emulate human awareness of event severity.
\end{remark}


\section{Risk-aware control}\label{sec:synthesis}

In this section, we first formulate the control synthesis problem with the proposed risk metric. A natural approach is to maximize the rewards metric of co-safety specification and also keep the risk metric of safety specification in an acceptable range, as formulated below:
\begin{equation}
\max _{\bar{\pi}} \mathcal{V}_{\bar{\pi}}\left(\pmb{\omega} \models \psi_{cs}\right) \text { s.t. } \mathcal{R}_{\bar{\pi}}\left(\pmb{\omega} \models \psi_{s} \right) \leq r_{th},
\label{eq:problem}
\end{equation}
where $r_{th} \in \mathbb R$ is a constant threshold. We first reformulate this optimization problem into an LP problem. Afterward, we detail the control synthesis framework that integrates traffic rules and vehicle dynamics, followed by an evaluation of the synthesized policy in a simplified traffic scenario.
\subsection{Linear Programming Problem}
This problem can be formulated with discounted occupation measures defined over the product MDP, $\mathcal{P}$. Firstly, the discounted occupation measures with respect to product states, denoted by $\beta_{\bar{\pi}}(z)$, equates to the sum of the discounted occupation measure with respect to state-action pairs, denoted by  $\beta_{\bar{\pi}}(z, a)$, over action set $A$, i.e., $\beta_{\bar{\pi}}(z) = \sum_{a \in A} \beta_{\bar{\pi}}(z,a)$. Specifically, $\beta_{\bar{\pi}}(z, a)$ is defined as,
\begin{equation}
\textstyle    \beta_{\bar{\pi}}(z, a): =\mathbb{E}_{\bar{\pi}}\left[\sum_{t=0}^{\infty} \gamma^t \mathbf{1}\left(z_t = z, a_t = a\right)\right].
\end{equation} 
For any $z^{\prime}\!\in\!Z$, the occupation measure at state $z^{\prime}$ should satisfy the balance constraint to ensure the equivalence between the incoming transitions and the initial state distribution with the occupation measure at $z^{\prime}$,
\begin{equation}
 \beta_{\bar{\pi}}(z^{\prime}) = \sum_{z \in Z} \sum_{a \in A} \gamma \beta_{\bar{\pi}}(z, a) \hat{\mathbb{T}}(z, a, z^{\prime}) +
\mathbf{1}(z^{\prime} = z_0), 
\label{eq:balance}
\end{equation}
A solution to this balance equation gives a stationary policy,
\begin{equation}\label{eq:oc_policy}
\bar{\pi}(z, a)= \frac{\beta_{\bar{\pi}} (z, a)}{\sum_{a \in A} \beta_{\bar{\pi}} (z, a)} = \frac{\beta_{\bar{\pi}}(z, a)}{\beta_{\bar{\pi}} (z)}.
\end{equation}
For this policy, the reward metric for co-safety formula $\psi_{cs}$ is formulated with discounted occupation measures as, 
\begin{equation}\label{eq:oc_reward_metric}
\overline{\mathcal{V}}_{\bar{\pi}}(\lozenge G)\!=\!\sum_{z^{\prime} \in G} \sum_{z \in Z} \sum_{a \in A} \gamma \beta_{\bar{\pi}}(z, a) \hat{\mathbb{T}}(z, a, z^{\prime}) + \mathbf{1}(z_0 \in G).
\end{equation}
To sum up, the policy, reward metric, and risk metric are formulated with occupation measures in~\eqref{eq:oc_policy},~\eqref{eq:oc_reward_metric} and~\eqref{eq:oc_risk_metric}, respectively. Furthermore, the control synthesis problem in~\eqref{eq:problem} can be formulated as a linear programming problem using $\beta_{\bar{\pi}}(z,a)\!\geq\!0$ for all $(z, a)\!\in\!Z\!\times\!A$ as decision variables,
\begin{equation}
	\begin{aligned}
		&\max_{ \{ \beta_{\bar{\pi}}(z,a) \} } && \textstyle  \bar{\mathcal{V}}_{\bar{\pi}}(\lozenge G)  
		\\
		& \text{s.t.} && \textstyle \sum_{z \in Z} \sum_{a \in A} c(z) \beta_{\bar{\pi}}(z, a)  \leq r_{th}  \\ 
		&  && \beta_{\bar{\pi}}(z, a) \geq 0, \; \forall (z, a) \in Z \times A\\
		&  && \textstyle \beta_{\bar{\pi}}(z^{\prime}) =\eqref{eq:balance}, \; \forall z'\in Z
	\end{aligned}
	\label{eq:lp}
\end{equation} 
The optimal policy $\bar{\pi}$ can be extracted from the optimized occupation measures using Eq.~\eqref{eq:oc_policy}, and then $\bar{\pi}$ can convert to $\pi\!\in\!\Pi$ defined on $\mathcal{M}_v$. This LP problem can be solved using Gurobi~\cite{gurobi} solver. 

\begin{figure}[ht]
        \vspace{-2mm}
	\centering
	\includegraphics[width=0.45
	\textwidth]{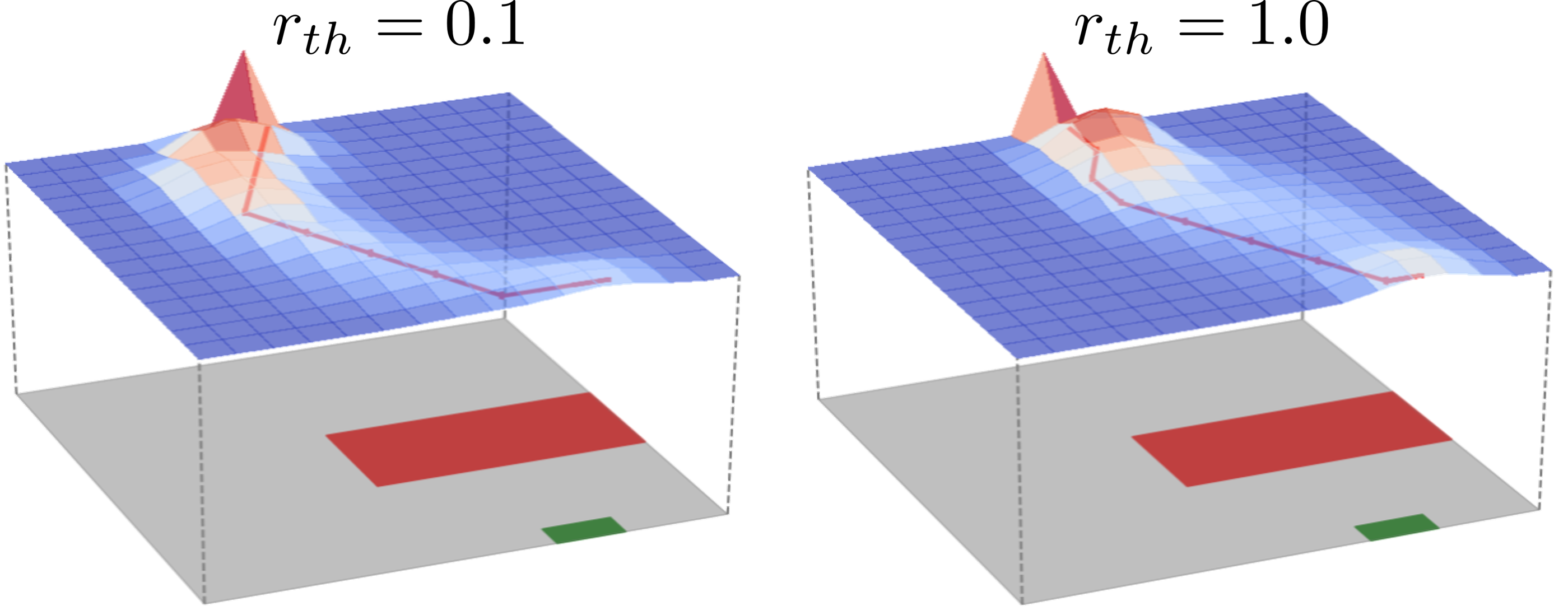}
	\caption{Risk field extracted from discounted occupation measures. The higher layer is the risk field, while the lower layer is a map of a reach-avoid problem, where the red rectangle is an obstacle and the green rectangle is the target. }
	\label{fig:risk_lp} \vspace{-3mm}
\end{figure}

To visualize the risk field calculated by the LP problem intuitively, we first consider a reach-avoid problem over the planar position space of a vehicle, excluding the dynamic environment and LTL specifications. Fig.~\ref{fig:risk_lp} displays the risk field comprising discounted occupation measures obtained by solving the LP problem in Eq.~\eqref{eq:lp}. The red solid line represents the optimal path determined by the occupation measures. As shown in the figure, we can see that the risk field obtained by the occupation measure aligns with the pattern of the heuristically designed risk field in~\cite{kolekar2020human}, revealing the underlying equivalence between the proposed metric and DRF. In addition, the different risk thresholds $r_{th}$ in Eq.~\eqref{eq:lp} will alter the shape of the risk field and further tune the conservatism of the generated policy.

\subsection{Implementation Framework} 
\begin{figure}[thb]
\vspace{2mm}
 	\centering
 	\includegraphics[width=0.46
 	\textwidth]{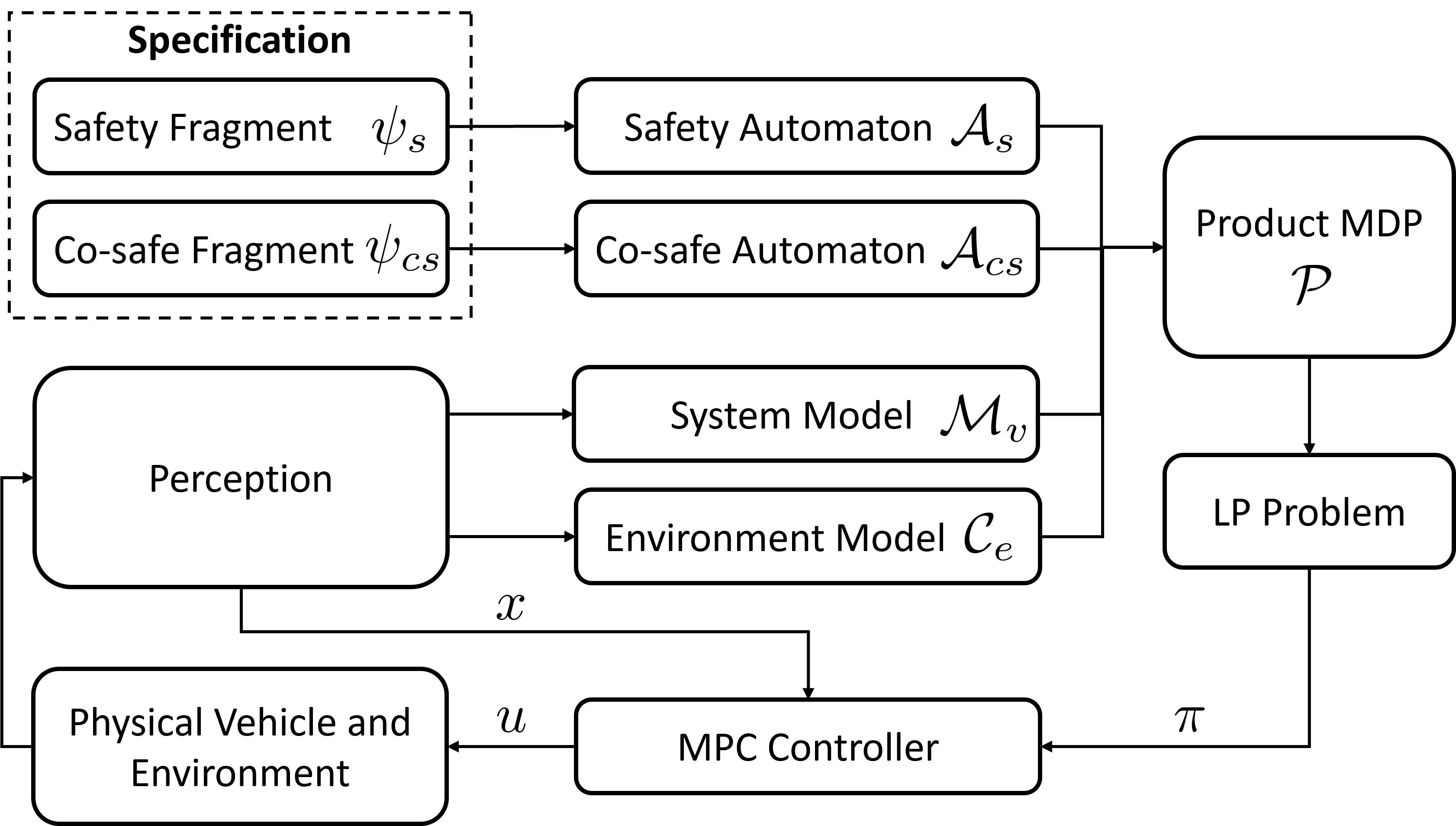}
 	\caption{The overall framework of control synthesis. 
    }
 	\label{fig:framework}
 \end{figure}

The above discussion is based on an MDP model of vehicles. However, to deploy this approach in a real car, we need to abstract an MDP model from vehicle dynamics and determine how to implement high-level policies obtained by the LP problem. The overall framework is shown in Fig.~\ref{fig:framework}.

 \noindent \textbf{Vehicle dynamics:} In the experiments, we consider the dynamics of vehicles with a discrete-time bicycle model. Specifically, the system state consists of the planar position, heading angle, and linear speed of the vehicle, denoted by $[p_x, p_y]$, $\theta$, and $v$, respectively. The control inputs to the vehicle are steering angle and acceleration, denoted by $\phi$ and $a$. Consequently, the system state is $\mathbf{x}(t)\!=\![p_x(t), p_y(t), \theta(t), v(t)]^T \!\in \! \mathbb{X} \!\subset\!\mathbb{R}^4$, and the system input is $\mathbf{u}(t) = [\phi(t), a(t)]^T\!\in\!\mathbb{U}\!\subset\!\mathbb{R}^2$, where $\mathbb{X}$ and $\mathbb{U}$ are state space and input space, respectively. The kinematic model of the vehicle is a bicycle model $\mathbf{x}(t+1)\!=\!f(\mathbf{x}(t), \mathbf{u}(t))$, and its details can be found in~\cite{zhang2024intention}.
 To capture the environmental uncertainty, a zero-mean Gaussian-distributed disturbance vector, $\mathbf{w} \sim \mathcal{N}(0, \Sigma)$, is considered with the variance $\Sigma \in \mathbb{R}^{4\times4}$. The vehicle model with uncertainty is denoted by $\mathbf{x}(t+1)\!=\!f(\mathbf{x}(t), \mathbf{u}(t)) + \mathbf{w}(t)$.


 \noindent \textbf{From continuous dynamics to MDP:} To abstract $\mathcal{M}_v$ from the vehicle dynamics model with uncertainty, we first discretize the state space $\mathbb{X}$ by gridding, where the finite state set $S$ comprises the centers of the grid cells. For simplicity, this implementation considers only the planar positions of vehicles in $\mathcal{M}_v$. Similarly, the continuous input space $\mathbb{U}$ is discretized into a finite action set $A$ using appropriate interface functions. In this implementation, the transition probabilities are predefined by experts, which can also be generated autonomously and formally~\cite{van2022temporal}. Additionally, the set of atomic propositions $\AP$ is determined by corresponding propositions in specifications, and the labeling function $\mathcal{L}$ can be extracted from the semantic maps in the perception module. Since perception is beyond the scope of this paper, the labels are manually assigned in this implementation.

 \noindent \textbf{Control refinement:} By solving the LP problem in Eq.~\eqref{eq:lp}, an optimal policy, $\pi$ for $\mathcal{M}_v$, is obtained such that the next action can be determined according to $\pi$. To track the reference state denoted by $\mathbf{x}_{r}\!\in\!\mathbb{X}$ corresponding to the next action, a model predictive controller (MPC) is designed to drive the dynamic model of vehicles, formulated as follows, 
 \begin{equation}
 	\begin{aligned}
 		& \textstyle \underset{\{\mathbf{u}_k\}_{k=0}^{N-1}}{\text{min}}
 		&& \textstyle \sum_{k=0}^{N-1} (\mathbf{x}_{k+1}\!-\!\mathbf{x}_{r})^\text{T} \mathbf{Q} (\mathbf{x}_{k+1}\!-\!\mathbf{x}_{r}) + \mathbf{u}_k^\text{T} \mathbf{R} \mathbf{u}_k \\
 		& \text{s.t.}
 		&& \mathbf{x}_{k+1} = f(\mathbf{x}_k, \mathbf{u}_k), \quad \mathbf{x}_0 = X_0,\\
 		& && \mathbf{x}_{k+1} \in \mathbb{X}, \quad \mathbf{u}_k \in \mathbb{U}, \quad k\!=\!0,\!\dots,\!N-1, 
 	\end{aligned}
 	\label{eq:mpc}
 \end{equation}
 where $N$ denotes a horizon number, $X_0$ is the current state, $\mathbf{Q}\!\in\!\mathbb{R}^{4\times4}$ and $\mathbf{R} \!\in\!\mathbb{R}^{2\times2}$ are diagonal weight matrices with positive weights. 

\subsection{Pedestrian Crossing Example} \label{sec:ped}
\begin{figure}[t]
\centering
\vspace{2mm}
    \includegraphics[width=0.95\linewidth]{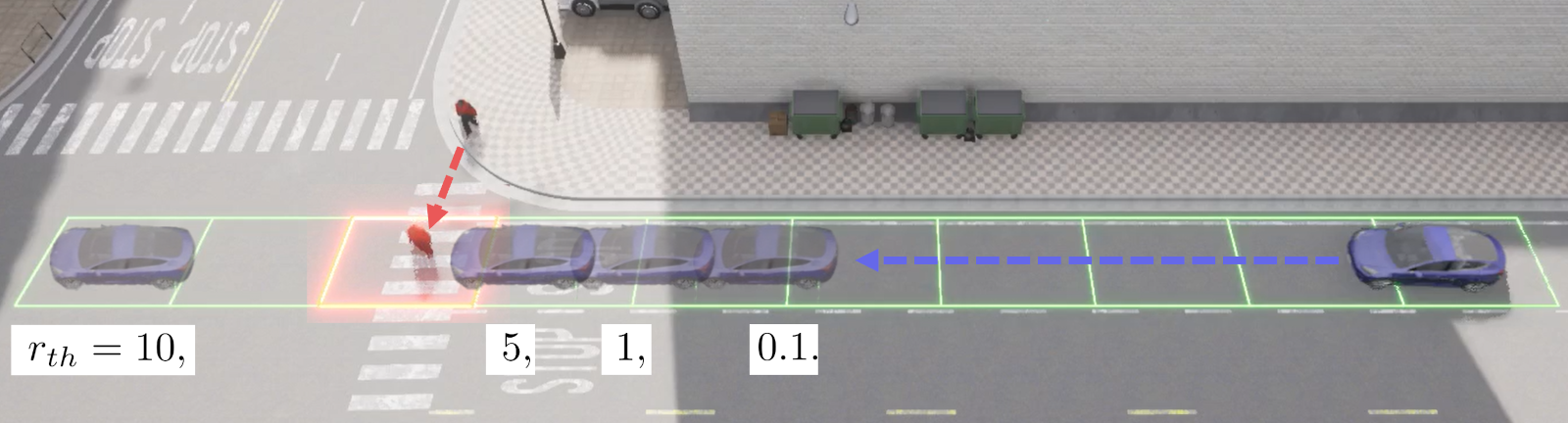} 
    \caption{Stopping points for different risk thresholds.}
    \label{fig:pedestrian}
\end{figure}

We first consider a simple scenario with a dynamic environment where a pedestrian is crossing at a crosswalk. The ego vehicle is expected to respond appropriately to the pedestrian's crossing behavior. To this end, we build an MC model to represent the uncertain behavior of the pedestrian. The specification of the ego vehicle is $\square(p\!\rightarrow\!\neg\! c) \wedge \lozenge(t)$, where $c$ and $t$ denote the crosswalk and target zone, respectively, and $p$ represents the presence of a pedestrian on the crosswalk. The cost of hitting a pedestrian is set to $8$. The ego vehicle is expected to autonomously decide whether to stop or proceed based on the real-time situation. To test risk sensitivity, we set different risk thresholds $r_{th}$ for the LP problem. The simulation results, depicted in Fig.~\ref{fig:pedestrian}, indicate that the ego vehicle will stop at a smaller clearance ahead of the crosswalk as the risk threshold increases, i.e., $r_{th} = 0.1, 1, 5$. However, when $r_{th} = 10$, the collision risk is ignored, resulting in the ego vehicle striking the pedestrian. This result demonstrates that the parameter $r_{th}$ can be used to tune the conservatism of the synthesized policy. 



However, tuning $r_{th}$ becomes an issue in more complex scenarios. The LP problem might be infeasible in risky states, requiring a larger $r_{th}$, but increasing $r_{th}$ leads to unreasonably aggressive behavior in less risky states. As a result, while the formulation in~\eqref{eq:lp} performs well in the pedestrian scenario, reasonable behavior cannot be achieved by tuning $r_{th}$ in complex scenarios such as unprotected turns in Fig.~\ref{fig:intersection}. To address this issue, we propose an improved method in the next section.

\section{Improved risk-aware control and experimental study}
Ideally, human drivers can actively minimize real-time risks while avoiding not only catastrophic outcomes but also overly conservative behavior. To achieve such a multi-objective balance, we introduce a soft risk threshold $r_s$ and a hard risk threshold $r_h$ to manage risk at different levels. This gives the improved control synthesis problem as follows,
\begin{equation}
	\begin{aligned}
		&\max_{ \{ \beta_{\bar{\pi}}(z,a) \} } && \textstyle  \bar{\mathcal{V}}_{\bar{\pi}}(\lozenge G) - K \xi
		\\
		& \text{s.t.} && \textstyle \sum_{z \in Z} \sum_{a \in A} c(z) \beta_{\bar{\pi}}(z, a)  \leq r_{s} + \xi,\\
		&   && \xi + r_s \leq r_{h}, \, \xi\geq0, \\
		&  && \beta_{\bar{\pi}}(z, a) \geq 0, \; \forall (z, a) \in Z \times A,\\
		&  &&\textstyle \beta_{\bar{\pi}}(z^{\prime}) = \eqref{eq:balance}, \; \forall z'\in Z,
	\end{aligned}
	\label{eq:improved_lp}
\end{equation} 
where $\xi$ denotes the relaxation variable for soft risk threshold $r_{s}$ and $K\!>\!0$ is a constant to balance the satisfaction of the co-safety formulas and risks of safety formulas. The risk metric is minimized when exceeding the soft threshold is unavoidable. In addition, $r_s$ filters out overly conservative behavior and $r_h$ ensures critical safety.


\color{black}
\subsection{ Experimental Study}
\label{sec:results}
We validate the improved framework in three scenarios, using default parameters: risk thresholds $r_s\!=\!1$ and $r_h\!=\!2$, and discounting factor $\gamma\!=\!0.8$. The video demonstration is accessible at link~\footnote{{\url{https://youtu.be/r5kEMW8oPQE}}} and the code is provided at link~\footnote{{\url{https://github.com/Miracle-qi/Risk_LTL_Planning}}}.


\noindent \textbf{1) Pedestrian crossing scenario:} The improved method can generate safe behavior with tunable conservatism under the same settings as in Sec.~\ref{sec:ped}, encompassing functionalities of the original one in Eq.~\eqref{eq:lp}. 

\begin{figure}[ht]
    \centering \vspace{-2mm}
    \includegraphics[width=0.47\textwidth]{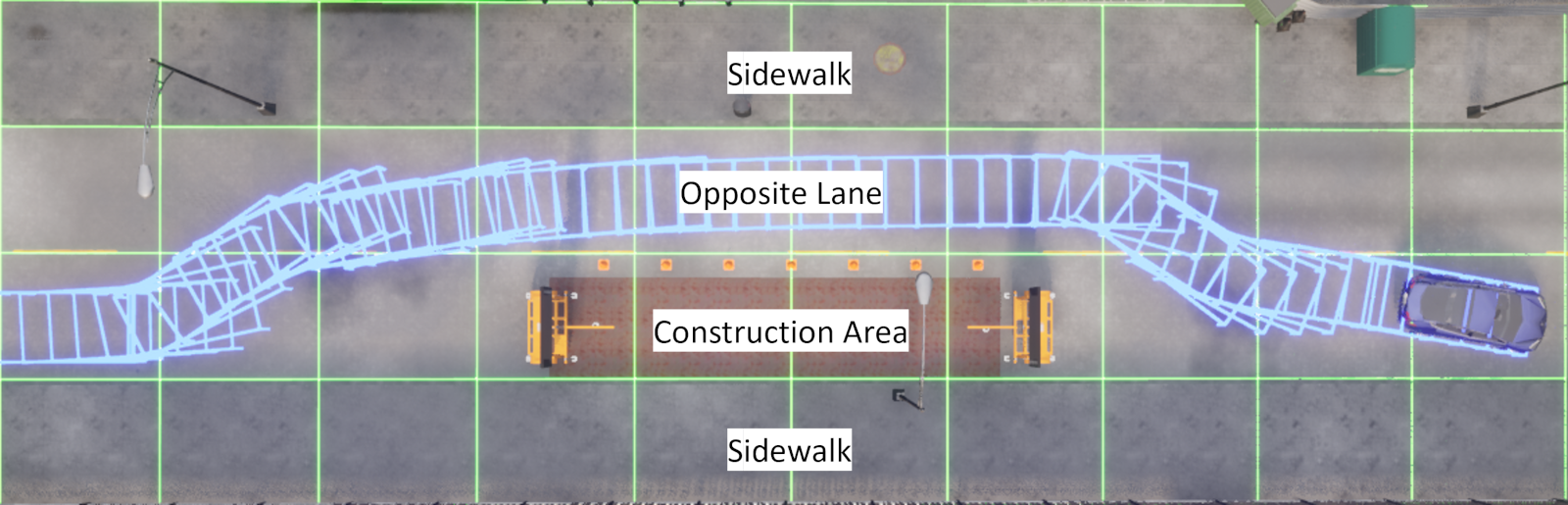}
    \caption{The generated minimal-violation trajectory. }
    \label{fig:construction} \vspace{-2mm}
\end{figure}

\noindent \textbf{2) Unexpected construction scenario:} Real-world traffic often contains unexpected situations that make the original specifications infeasible~\cite{wongpiromsarn2021minimum}, such as traffic participants violating rules or disruptions of traffic environments. We tested the proposed methods in such a scenario where an unexpected construction zone appears in the forward lane of the ego car, as illustrated in Fig.~\ref{fig:construction}. In this case, the ego vehicle must deviate from a part of the original safety specifications to reach its target. The specification of this scenario for the ego vehicle is $\square(\neg c \wedge \neg o \wedge \neg s) \wedge \lozenge(t)$, where $c$, $o$, $s$, and $t$ are labels for construction areas, opposite lanes, sidewalks, and target areas. We assign cost values of $5$, $3$, and $1$ to the construction area, sidewalk, and opposite lane, respectively. Figure~\ref{fig:construction} shows the generated trajectory, where the ego vehicle autonomously chooses to bypass the construction zone via the opposite lane with minimal violation and finally reaches the target. This result demonstrates that the improved control method effectively balances different risks and minimizes violations. This advantage arises from incorporating severity into LTL specifications. 

\noindent \textbf{3) Unprotected turn scenario:} We consider an intersection scenario that includes a traffic light and dynamic opponent vehicles, as depicted in Fig.~\ref{fig:intersection}. The ego vehicle (blue car) is expected to follow traffic signals and avoid both non-drivable areas and the opponent vehicles (red car). The specification has been introduced in the study case of Sec.~\ref{sec:ltl}. An MC model is designed to represent the switching of traffic lights and the movement of the opponent vehicle. For simplicity, we assume the opponent vehicle always moves forward. The collision risk with the opponent vehicle is evaluated by calculating the overlap between risk fields of vehicles, following the method described in~\cite{joo2023generalized}. With the policy synthesized by the LP problem, the ego vehicle will wait at the boundary of the intersection until the light turns green, adjust its decisions based on the real-time positions of the opponent vehicles, and finally reach the target position. The whole process is illustrated in Fig.~\ref{fig:intersection} and the supplementary video. This case accounts for traffic lights, non-drivable areas, and dynamic vehicles, which are typical factors in traffic. Thus, this method can handle complex situations and has the potential to scale to various scenarios.

\begin{figure}[tb] 
\vspace{1mm}
    \centering
    \includegraphics[width=0.46\textwidth]{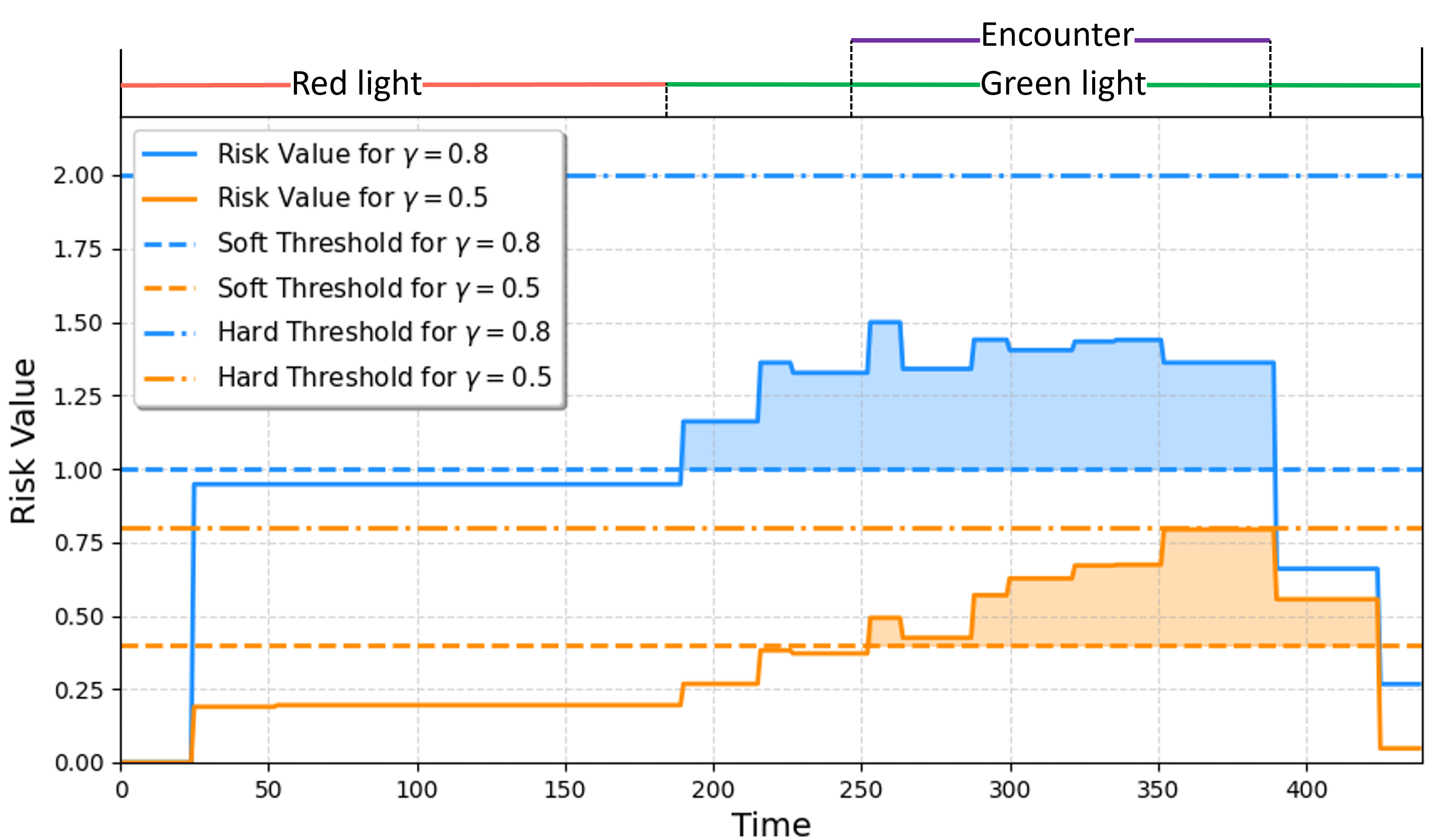}
    \caption{Risk curves under different discounting factors.}
    \label{fig:risk_profile}
\end{figure}

To examine the influence of discounting factors, the corresponding risk curves are shown in Fig.~\ref{fig:risk_profile}. While the generated behavior remains identical across different discounting factors, the risk curves in the process are different. Compared to $\gamma\!=\!0.8$, the discounting factor $\gamma\!=\!0.5$ places more focus on recent events, causing the risk curves of $\gamma\!=\!0.5$ to gradually increase as the two vehicles approach each other. In contrast, the risk curves for $\gamma\!=\!0.8$ remain flatter because the risks are perceived earlier. Since placing more focus on recent events lowers perceived risk values, the thresholds for lower discounting factors are tuned down accordingly.

Table~\ref{tb:computation} presents the maximum and average risk values with the default discounting and threshold, along with the state number and computation time for the LP problem in~\eqref{eq:lp}. The results show that the improved LP problem effectively manages different types of risk and maintains gross risk value within the predefined level throughout the execution process. However, computation time increases significantly as the number of states grows, making finer abstractions for realistic traffic scenarios computationally expensive.


\begin{table}[ht]
\centering 
\begin{tabular}{c|c|c|c}
\hline
& \textbf{Scenario 1} & \textbf{Scenario 2} & \textbf{Scenario 3} \\ 
\hline
Maximal risk &0.8 & 1.46 & 1.51 \\
Average risk &0.68 & 0.97 & 1.01 \\ \hline
State number & 1440 &  720 & 2880\\
Computation time/ms & 20-30 & 15-20 & 50-60 \\ \hline
\end{tabular}
\caption{Risk value and computation time of three scenarios.}
\label{tb:computation} 
\end{table}

\section{Conclusion} \label{sec:con}

In this paper, we extend a human-like risk metric to account for complex traffic scenarios by integrating LTL specifications. At the same time, the proposed metric makes temporal logic more practical for traffic tasks by incorporating human-like awareness. Furthermore, a risk-aware control method is improved to balance different types of risks in a human-like manner. Extensive simulation results validate the effectiveness and potential of this framework. To better handle realistic traffic, future work should incorporate strategic interactions between vehicles, reduce computational costs for finer abstractions, and learn parameters from real-road data instead of manual configuration.


\bibliographystyle{ieeetr}        
\bibliography{MyBib}    

@article{kolekar2020human,
  title={Human-like driving behaviour emerges from a risk-based driver model},
  author={Kolekar, Sarvesh and de Winter, Joost and Abbink, David},
  journal={Nature Communications},
  volume={11},
  number={1},
  pages={4850},
  year={2020},
  publisher={Nature Publishing Group UK London}
}

@inproceedings{qi2023automated,
  title={Automated Formation Control Synthesis from Temporal Logic Specifications},
  author={Qi, Shuhao and Zhang, Zengjie and Haesaert, Sofie and Sun, Zhiyong},
  booktitle={IEEE Conference on Decision and Control (CDC)},
  pages={5165--5170},
  year={2023},
  organization={IEEE}
}

@book{belta2017formal,
  title={Formal methods for discrete-time dynamical systems},
  author={Belta, Calin and Yordanov, Boyan and Gol, Ebru Aydin},
  volume={89},
  year={2017},
  publisher={Springer}
}

@article{joo2023generalized,
  title={A generalized driving risk assessment on high-speed highways using field theory},
  author={Joo, Yang-Jun and Kim, Eui-Jin and Kim, Dong-Kyu and Park, Peter Y},
  journal={Analytic methods in accident research},
  volume={40},
  pages={100303},
  year={2023},
  publisher={Elsevier}
}

@article{van2022temporal,
  title={Temporal logic control of nonlinear stochastic systems using a piecewise-affine abstraction},
  author={van Huijgevoort, Birgit C and Weiland, Siep and Haesaert, Sofie},
  journal={IEEE Control Systems Letters},
  volume={7},
  year={2022},
  publisher={IEEE}
}

@article{ulusoy2014incremental,
  title={Incremental controller synthesis in probabilistic environments with temporal logic constraints},
  author={Ulusoy, Alphan and Wongpiromsarn, Tichakorn and Belta, Calin},
  journal={The International Journal of Robotics Research},
  volume={33},
  number={8},
  pages={1130--1144},
  year={2014},
  publisher={SAGE Publications Sage UK: London, England}
}

@inproceedings{majumdar2020should,
  title={How should a robot assess risk? towards an axiomatic theory of risk in robotics},
  author={Majumdar, Anirudha and Pavone, Marco},
  booktitle={Robotics Research: The 18th International Symposium ISRR},
  pages={75--84},
  year={2020},
  organization={Springer}
}

@article{lindemann2021reactive,
  title={Reactive and risk-aware control for signal temporal logic},
  author={Lindemann, Lars and Pappas, George J and Dimarogonas, Dimos V},
  journal={IEEE Transactions on Automatic Control},
  volume={67},
  number={10},
  pages={5262--5277},
  year={2021},
  publisher={IEEE}
}

@article{wongpiromsarn2023formal,
  title={Formal Methods for Autonomous Systems},
  author={Wongpiromsarn, Tichakorn and Ghasemi, Mahsa and Cubuktepe, Murat and Bakirtzis, Georgios and Carr, Steven and Karabag, Mustafa O and Neary, Cyrus and Gohari, Parham and Topcu, Ufuk},
  journal={Foundations and Trends in Systems and Control},
  volume = 10,
number = {3-4}, 
pages= {180-407},
  year={2023}
}

@inproceedings{wongpiromsarn2021minimum,
  title={Minimum-violation planning for autonomous systems: Theoretical and practical considerations},
  author={Wongpiromsarn, Tichakorn and Slutsky, Konstantin and Frazzoli, Emilio and Topcu, Ufuk},
  booktitle={2021 American Control Conference (ACC)},
  pages={4866--4872},
  year={2021},
  organization={IEEE}
}

@book{altman2021constrained,
  title={Constrained Markov decision processes},
  author={Altman, Eitan},
  year={2021},
  publisher={Routledge}
}

@article{althoff2025no,
  title={No More Traffic Tickets: A Tutorial to Ensure Traffic-Rule Compliance of Automated Vehicles},
  author={Althoff, Matthias and Maierhofer, Sebastian and W{\"u}rsching, Gerald and Lin, Yuanfei and Lercher, Florian and Stolz, Roland},
  journal={Proceedings of the IEEE},
  year={2025},
  publisher={IEEE}
}

@inproceedings{haesaert2021formal,
  title={Formal multi-objective synthesis of continuous-state MDPs},
  author={Haesaert, Sofie and Nilsson, Petter and Soudjani, Sadegh},
  booktitle={2021 American Control Conference (ACC)},
  pages={3428--3433},
  year={2021},
  organization={IEEE}
}

@article{mehdipour2023formal,
  title={Formal methods to comply with rules of the road in autonomous driving: State of the art and grand challenges},
  author={Mehdipour, Noushin and Althoff, Matthias and Tebbens, Radboud Duintjer and Belta, Calin},
  journal={Automatica},
  volume={152},
  year={2023},
  publisher={Elsevier}
}

@inproceedings{trevizan2017occupation,
  title={Occupation measure heuristics for probabilistic planning},
  author={Trevizan, Felipe and Thi{\'e}baux, Sylvie and Haslum, Patrik},
  booktitle={Proceedings of the International Conference on Automated Planning and Scheduling},
  volume={27},
  year={2017}
}

@inproceedings{artale2023complexity,
  title={Complexity of safety and cosafety fragments of linear temporal logic},
  author={Artale, Alessandro and Geatti, Luca and Gigante, Nicola and Mazzullo, Andrea and Montanari, Angelo},
  booktitle={Proceedings of the AAAI Conference on Artificial Intelligence},
  volume={37},
  pages={6236--6244},
  year={2023}
}

@inproceedings{henriksen1995mona,
  title={Mona: Monadic second-order logic in practice},
  author={Henriksen, Jesper G and Jensen, Jakob and J{\o}rgensen, Michael and Klarlund, Nils and Paige, Robert and Rauhe, Theis and Sandholm, Anders},
  booktitle={Tools and Algorithms for the Construction and Analysis of Systems: First International Workshop, TACAS'95 Aarhus, Denmark, May 19--20, 1995 Selected Papers 1},
  pages={89--110},
  year={1995},
  organization={Springer}
}

@misc{gurobi,
  author = {{Gurobi LLC}},
  title = {{Gurobi Optimizer Reference Manual}},
  year = 2024,
  url = "https://www.gurobi.com"
}

@inproceedings{dosovitskiy2017carla,
  title={{CARLA}: An open urban driving simulator},
  author={Dosovitskiy, Alexey and Ros, German and Codevilla, Felipe and Lopez, Antonio and Koltun, Vladlen},
  booktitle={Conference on robot learning},
  pages={1--16},
  year={2017},
  organization={PMLR}
}

@article{ahmed2020rationality,
  title={Rationality and future discounting},
  author={Ahmed, Arif},
  journal={Topoi},
  volume={39},
  number={2},
  pages={245--256},
  year={2020},
  publisher={Springer}
}

@article{geisslinger2021autonomous,
  title={Autonomous driving ethics: From trolley problem to ethics of risk},
  author={Geisslinger, Maximilian and Poszler, Franziska and Betz, Johannes and L{\"u}tge, Christoph and Lienkamp, Markus},
  journal={Philosophy \& Technology},
  volume={34},
  number={4},
  pages={1033--1055},
  year={2021},
  publisher={Springer}
}

@article{guo2018probabilistic,
  title={Probabilistic motion planning under temporal tasks and soft constraints},
  author={Guo, Meng and Zavlanos, Michael M},
  journal={IEEE Transactions on Automatic Control},
  volume={63},
  number={12},
  pages={4051--4066},
  year={2018},
  publisher={IEEE}
}

@inproceedings{lindemann2021stl,
  title={{STL} robustness risk over discrete-time stochastic processes},
  author={Lindemann, Lars and Matni, Nikolai and Pappas, George J},
  booktitle={2021 60th IEEE Conference on Decision and Control (CDC)},
  pages={1329--1335},
  year={2021},
  organization={IEEE}
}

@book{kochenderfer2015decision,
  title={Decision making under uncertainty: theory and application},
  author={Kochenderfer, Mykel J},
  year={2015},
  publisher={MIT press}
}

@article{wang2022social,
  title={Social interactions for autonomous driving: A review and perspectives},
  author={Wang, Wenshuo and Wang, Letian and Zhang, Chengyuan and Liu, Changliu and Sun, Lijun and others},
  journal={Foundations and Trends in Robotics},
  volume={10},
  number={3-4},
  pages={198--376},
  year={2022},
  publisher={Now Publishers, Inc.}
}

@article{zhang2024intention,
  title={Intention-Aware Control Based on Belief-Space Specifications and Stochastic Expansion},
  author={Zhang, Zengjie and Sun, Zhiyong and Haesaert, Sofie},
  journal={IEEE Transactions on Intelligent Vehicles},
  year={2024},
  publisher={IEEE}
}


\end{document}